\documentclass[journal]{IEEEtran}
\usepackage[utf8]{inputenc}
\usepackage[T1]{fontenc}
\usepackage{lmodern} % Enhanced font
\usepackage{amsmath, amssymb} % Math packages
\usepackage{graphicx} % For figures
\usepackage{hyperref} % For hyperlinks
\usepackage{cite} % For IEEE-style numeric citations
\usepackage{times}
\usepackage{adjustbox} % For adjustbox environment
\usepackage{tikz}
\usepackage{pgfplots}
\usepackage{rotating}
\usepackage[american,siunitx]{circuitikz}

\usetikzlibrary{shapes.geometric, arrows, arrows.meta, positioning,  3d, calc, fit}

\title{Design and Implementation of a Multi-Purpose Low-Cost Hall-Effect Sensor Glove for Sign Language Recognition}

\author{Dinanath Padhya, Jenish Pant, Krishna Acharya, Sajen Maharjan, Sudip Kumar Thakur\\
Thapathali Campus, Institute of Engineering, Kathmandu, Nepal\\
(dinanath, jenish.078bei018, krishna, sajen.078bei048, sudip.078bei044)@tcioe.edu.np%
\thanks{Corresponding author: Dinanath Padhya}%
}

\begin{document}

\maketitle

\begin{abstract}
    Despite the prevalence of severe hearing loss affecting over 430 million people globally, access to sign language interpretation remains critically scarce, particularly in low-resource settings like Nepal. Assistive technologies divide into two flawed categories: prohibitively expensive commercial gloves (often exceeding \$3,000) or fragile research prototypes reliant on flex sensors that degrade rapidly under mechanical stress. This paper introduces a robust, cost-effective sign language recognition system tailored for the Nepali Sign Language (NSL) community. Departing from traditional resistive sensing, we implement a non-contact Hall-effect architecture that correlates magnetic field intensity with finger flexion, eliminating mechanical wear and signal drift. The system integrates 14 sensor nodes across the DIP, PIP, and MCP joints, augmented by an MPU6050 IMU for wrist orientation. An embedded Multi-Layer Perceptron, executed locally on an Arduino Mega, performs gesture classification, negating the need for cloud dependencies. With a Bill of Materials between \$80 and \$100, this solution is approximately 30 times more affordable than market alternatives. Validation trials across five subjects yielded 96\% accuracy on a fundamental NSL vocabulary. As a proof-of-concept demonstration, this work validates the approach using 11 common NSL words, establishing a foundation for future vocabulary expansion. Stress testing confirmed that the Hall-effect configuration maintains signal fidelity over repeated cycles where traditional sensors fail. This study demonstrates that high-precision recognition is achievable through strategic engineering rather than premium components, offering a scalable pathway for deployment in Nepal's deaf schools.
\end{abstract}

\textbf{Index Terms-} Sign language recognition, Hall-effect sensors, wearable sensors, embedded neural networks, low-cost design
\section{Introduction}
Hearing loss imposes a profound barrier to social integration, employment, and education for over 430 million people worldwide. In developing nations like Nepal, this exclusion is exacerbated by a chronic shortage of trained interpreters and educational resources. While the digital divide in urban centers is narrowing, rural access to sign language education remains virtually non-existent.

Technological interventions have attempted to bridge this gap, yet current paradigms suffer from inherent structural flaws. Vision-based systems, while promising, demand significant computational resources and struggle with environmental variables such as poor lighting and occlusion. Conversely, wearable "smart gloves" offer a direct interface for gesture capture but are stifled by a dichotomy of cost and durability. Commercial solutions, such as the Bright Sign Glove, present a financial barrier so high (often $>\$3,000$) that they are irrelevant to the Global South. Meanwhile, the research landscape is saturated with prototypes using flex sensors \cite{prasetijo_rancang_2018}. These resistive strips are fundamentally ill-suited for long-term use; the act of bending causes material fatigue, leading to inevitable signal drift and device failure.

This paper proposes a divergence from standard resistive sensing. We present the design and implementation of a durable, low-cost recognition glove specifically optimized for Nepali Sign Language (NSL). Our primary contribution is the architectural shift to non-contact Hall-effect sensors. By measuring magnetic flux rather than material deformation, we decouple the sensing mechanism from mechanical wear, ensuring long-term reliability at a fraction of the cost. The following sections detail the system's electromechanical design, the on-board Artificial Neural Network (ANN) for static gesture classification, and validation results achieving 96\% accuracy, proving that accessibility need not come at the expense of precision.

\section{Related Work}
Engineers typically design recognition gloves by fusing flex sensors with Inertial Measurement Units (IMUs) \cite{burhani_sign_2023}. This combination tracks finger bending and hand orientation. When paired with Artificial Neural Networks (ANNs), these systems achieve high accuracy for static gestures \cite{burhani_sign_2023, b_survey_2022, khan_deep_2025}. Results often exceed 98-99\% for languages like American Sign Language (ASL) \cite{burhani_sign_2023}.

Flex sensors introduce a critical mechanical weakness. They operate as variable resistors. Bending degrades the resistive ink over time \cite{galvan-ruiz_perspective_2020, amin_comparative_2022}. Researchers have tested alternatives to address this flaw. These include piezoelectric elements \cite{galvan-ruiz_perspective_2020}, fiber optics \cite{prasetijo_rancang_2018}, knitted strain sensors \cite{ahmed_review_2018}, and auxetic-interlaced yarn sensors \cite{khan_deep_2025}. This search for new materials confirms that standard resistive sensors lack the necessary durability for daily use.

Cost presents a second barrier. Commercial hardware prices exclude most of the deaf community. Recent studies focus on accessibility. Prototyping with capacitive sensors and optimized electronics lowers expenses \cite{nunez-marcos_survey_2022, n_review_2024}. O'Connor et al. \cite{n_review_2024} and others \cite{amin_comparative_2022} proved that sub-\$100 designs are feasible. Our project targets this specific price benchmark.

Linguistic bias also limits current technology. Most research focuses on American Sign Language (ASL) \cite{burhani_sign_2023, prasetijo_rancang_2018}. Regional languages receive little attention. Studies exist for Polish \cite{galvan-ruiz_perspective_2020}, Korean \cite{ahmed_review_2018}, and Indian sign languages \cite{adeyanju_machine_2021}. Nepali Sign Language (NSL) remains largely undocumented in hardware studies.

We address these three gaps simultaneously. We use durable Hall-effect sensors to solve the durability problem. These sensors are robust, provide high precision, and resist mechanical wear. Our design uses inexpensive sensors and an efficient ANN model \cite{burhani_sign_2023, khan_deep_2025} to meet the sub-\$100 price point established by O'Connor et al. \cite{n_review_2024}. We validate the system specifically for the NSL community, providing a solution for a previously overlooked population.

\section{Methodology}
\subsection{System Architecture}
\begin{figure}[htbp]
    \centerline{\includegraphics[width=0.48\textwidth]{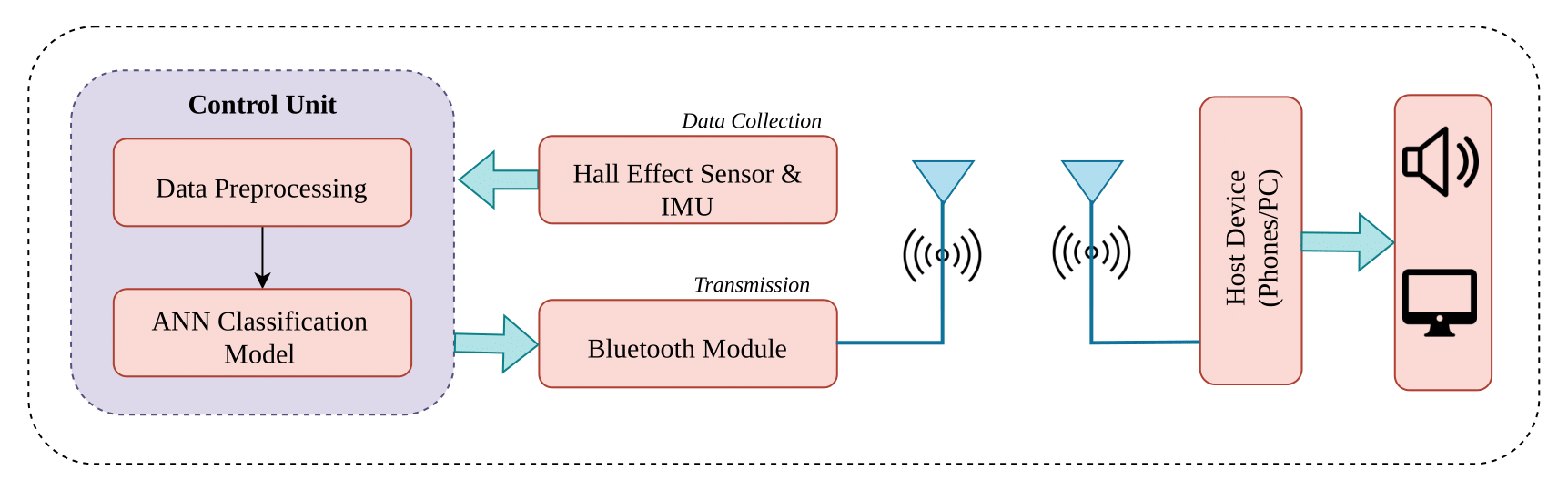}}
    \caption{System architecture showing data flow from sensor acquisition to speech output}
    \label{fig:system_architecture}
\end{figure}
The architecture shown in \ref{fig:system_architecture} operates as a four-stage pipeline starting with the Data Collection module, where Hall-effect sensors and an IMU capture finger flexion and wrist orientation. This data moves to the Control Unit, which executes Data Preprocessing and runs the embedded ANN Classification Model to identify the static gesture. The system then passes the classification index to the Transmission stage, which broadcasts the data via a Bluetooth Module. Finally, a Host Device such as a smartphone or PC receives the signal to execute the corresponding text-to-speech or control function.
\subsection{3D Structural Design}
\begin{figure}[htbp]
    \centering
    \includegraphics[width=0.40\textwidth]{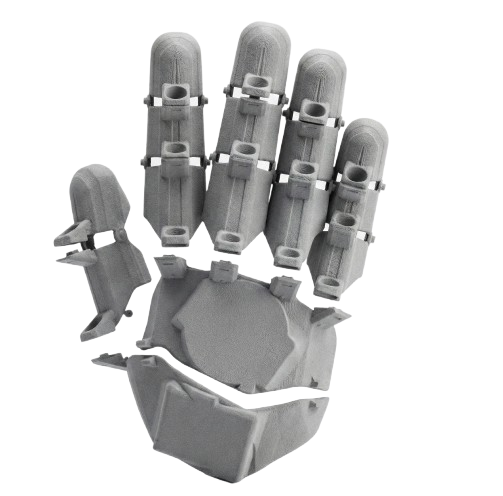}
    \caption{3D CAD model showing the glove's structural design and sensor mounting points}
    \label{fig:3d_model}
\end{figure}

The mechanical foundation of the system is a custom-designed 3D-printed exoskeleton, engineered to provide rigid and repeatable sensor placement while accommodating the natural kinematics of the human hand. The complete structural design is shown in Figure~\ref{fig:3d_model}.

The design employs a segmented architecture with individual ``sleeves'' fitted to each phalanx of the fingers. These sleeves are dimensioned based on anthropometric data from globally reported finger measurements, ensuring compatibility across users with varying hand sizes. Each segment is designed to house a Hall-effect sensor and its corresponding magnet while maintaining minimal bulk and weight.

The exoskeleton's modular construction allows for independent fabrication and assembly of each finger module, facilitating repairs and customization. Strategic placement of mounting points ensures that the magnetic field gradient remains consistent across the full range of joint motion. The backplate of the structure provides a stable platform for mounting the processing electronics, IMU, and power system, while also serving as the primary attachment point to the user's hand via adjustable straps. The current proof-of-concept (\ref{fig:prototype}) uses an Arduino Mega for rapid prototyping. Production versions could integrate a custom-designed ASIC to reduce size, power consumption, and cost.

\subsection{Anatomical Alignment and Design Validation}
The glove is engineered to capture the flexion--extension behaviour of all five fingers together with wrist orientation, reflecting the fundamental degrees of freedom (DoF) that encode Nepali Sign Language postures. Hand anthropometry and kinematic modelling informed every structural decision, ensuring that sensor placement maps consistently onto anatomical landmarks.

\textbf{Joint targeting.} The sensing layout covers the Distal Interphalangeal (DIP), Proximal Interphalangeal (PIP), and Metacarpophalangeal (MCP) joints. Each joint is fitted with a dedicated Hall-effect sensor positioned to track flexion angles throughout the full range of motion.

\begin{figure}[htbp]
    \centering
    \includegraphics[width=0.48\textwidth]{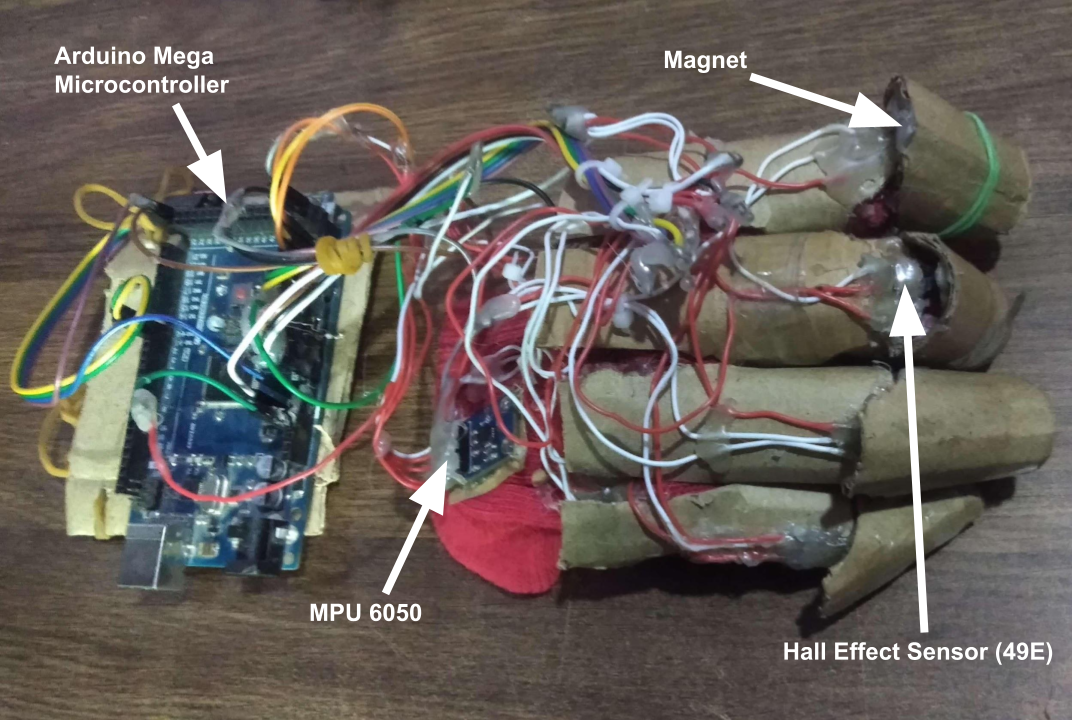}
    \caption{Early proof-of-concept prototype using cardboard chassis for initial sensor validation and testing}
    \label{fig:prototype}
\end{figure}

\textbf{Hybrid fulcrum strategy.} Early prototypes (Figure~\ref{fig:prototype}) demonstrated that rigidizing the MCP region suppresses the joint's intrinsic abduction/adduction capability, creating discomfort and skewing measurements. The final design therefore combines rigid fixtures on the DIP/PIP segments, which are primarily hinge joints, with flexible, rubber-band based anchors around the knuckles. This arrangement keeps each magnet, sensor pair co-linear during flexion while allowing the lateral compliance necessary for natural motion and long-term wearability. However, this flexibility introduces a trade-off: the rubber-band anchors permit minor positional variations that can introduce small signal fluctuations, particularly during rapid hand movements. These fluctuations are mitigated by the neural network's pattern recognition capability, which learns to filter such mechanical noise.
\subsection{Sensor Integration: Hall-Effect Mechanism}
\begin{figure}[htbp]
    \centering
    \includegraphics[width=0.40\textwidth]{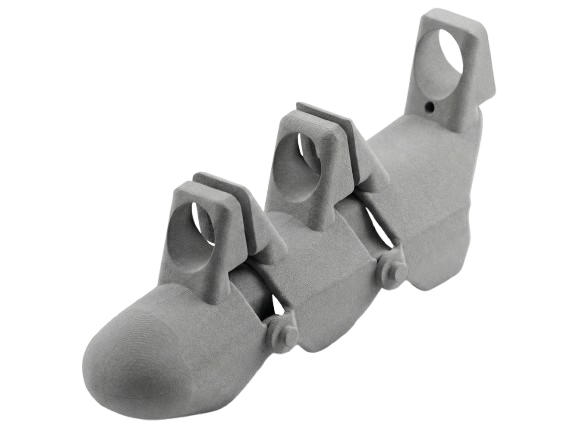}
    \caption{Hall-effect sensor and magnet mounting mechanism on a finger joint}
    \label{fig:sensor_attachment}
\end{figure}

The core transduction mechanism employs the 49E linear Hall-effect sensor, an analog IC selected for its ability to output a voltage proportional to magnetic flux density. Each monitored joint is equipped with a proximal sensor and a distal N35 neodymium magnet (10~mm diameter, 5~mm thickness). This pairing is compact enough to integrate within the 3D-printed phalanx segments while generating a magnetic field gradient strong enough to be read clearly over background noise (Figure~\ref{fig:sensor_attachment}).

\subsubsection{Magnetic Actuation Strategy.} The magnet-sensor configuration is designed such that flexion translates the magnet away from the sensing die. This results in a monotonic decay of magnetic flux, simplifying downstream signal processing as every joint angle correlates to a distinct voltage signature.

\subsubsection{Response Characterization.} Calibration is anchored by two physical boundary conditions: full extension (maximum flux/voltage) and full flexion (minimum flux). While the 49E sensor is nominally linear, the magnetic field intensity follows an inverse-cube law relative to distance. Consequently, the voltage-angle relationship exhibits sigmoidal behavior, saturating at extreme proximity and tapering off at maximum flexion (Figure~\ref{fig:raw_data}). Rather than using computationally expensive look-up tables to linearize this data on the microcontroller, we feed the normalized raw values directly into the neural network. The ANN is inherently capable of mapping these non-linear inputs to gesture classes, absorbing the sensor's geometric non-linearities into the model weights.

\subsubsection{Voltage response modelling.} 
\begin{figure}[htbp]
    \centering
    \includegraphics[width=0.48\textwidth]{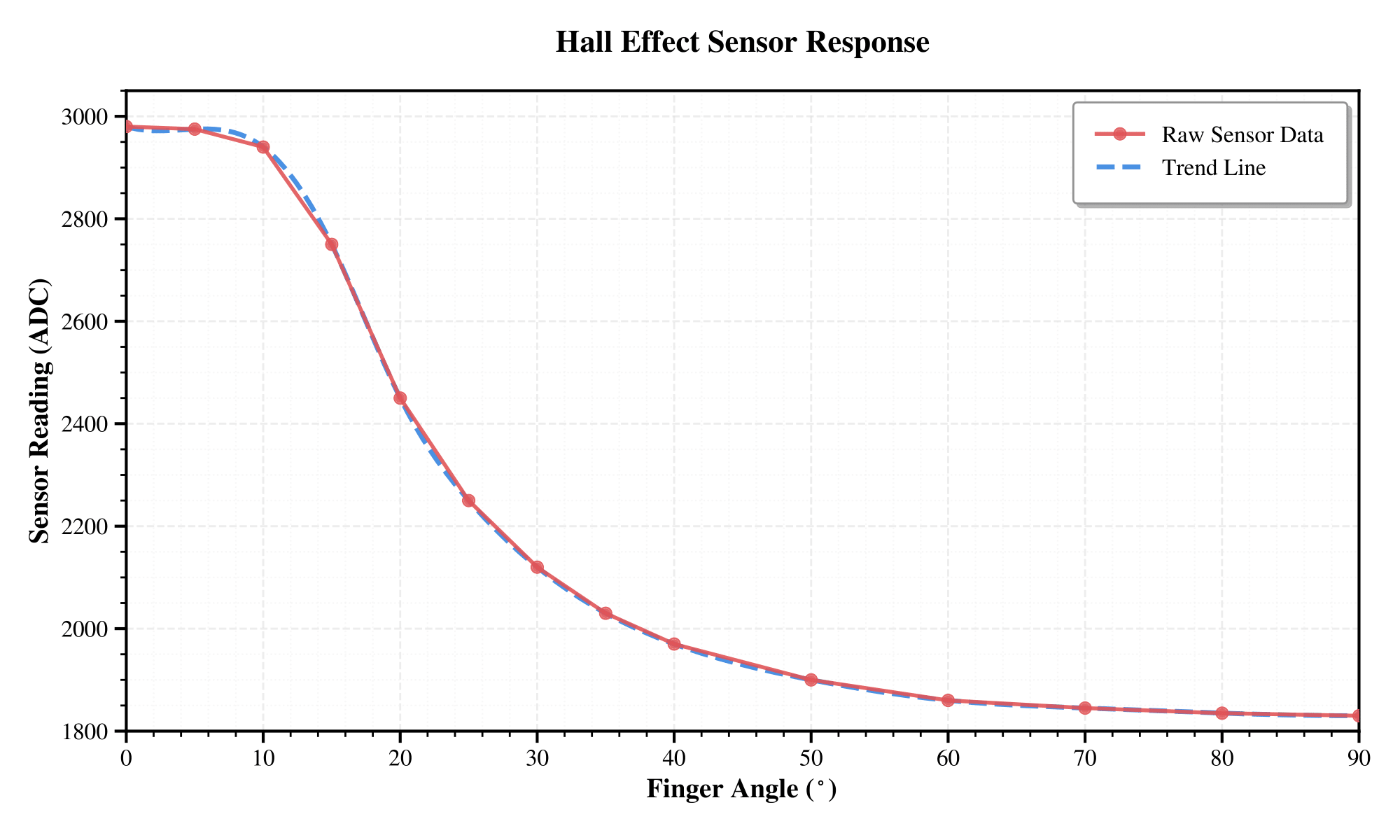}
    \caption{Nonlinear Hall-effect sensor response showing sigmoidal saturation at extreme flexion angles}
    \label{fig:raw_data}
\end{figure}

Although the manufacturer characterises the 49E at 5~V, the glove operates all electronics at 3.3~V for compatibility with the Arduino Mega. Because the sensor is ratiometric, the nominal response follows
\begin{equation}
    V_{\text{out}} = 0.5 V_{cc} + kB
\end{equation}
Extrapolating to the lower rail yields
\begin{equation}
    V_{\text{out}} \approx 1.6 \times 10^{-6} B + 1.65
    \label{eq:hall_voltage}
\end{equation}
keeping the analog signal within the Arduino's ADC span and removing the need for level shifters or voltage dividers. However, this linear approximation holds only within the sensor's mid-range operating region. As shown in Figure~\ref{fig:raw_data}, the actual voltage-angle relationship exhibits sigmoidal saturation behavior at the extremes of joint flexion, where the magnetic field becomes either too weak (fully flexed) or saturates the sensor (fully extended).

\subsubsection{Magnetic field interference mitigation.} With 14 N35 neodymium magnets in close proximity on the glove, potential cross-talk between adjacent sensors is a concern. However, the inverse-cube law of magnetic field decay ($B \propto 1/r^3$) provides natural isolation. At typical inter-finger distances ($>$2~cm), the field strength from adjacent magnets falls below the noise floor of neighboring sensors. Additionally, each magnet-sensor pair operates in a differential configuration where the sensor primarily responds to the local gradient, further reducing susceptibility to distant field sources.

\subsubsection{Sensor data normalization.} The raw analog readings from the Hall-effect sensors are normalized to the $[0,1]$ range before being fed to the neural network. This normalization ensures consistent input scaling across different sensors and users. Critically, the ANN learns to map these complex, nonlinear sensor patterns, including the saturation regions, directly to gesture classes, effectively correcting for the non-ideal sensor behavior without requiring explicit angle calculation or piecewise calibration.
\subsection{Electronic Circuit Architecture}
\begin{figure}[htbp]
    \centering
    \includegraphics[width=0.48\textwidth]{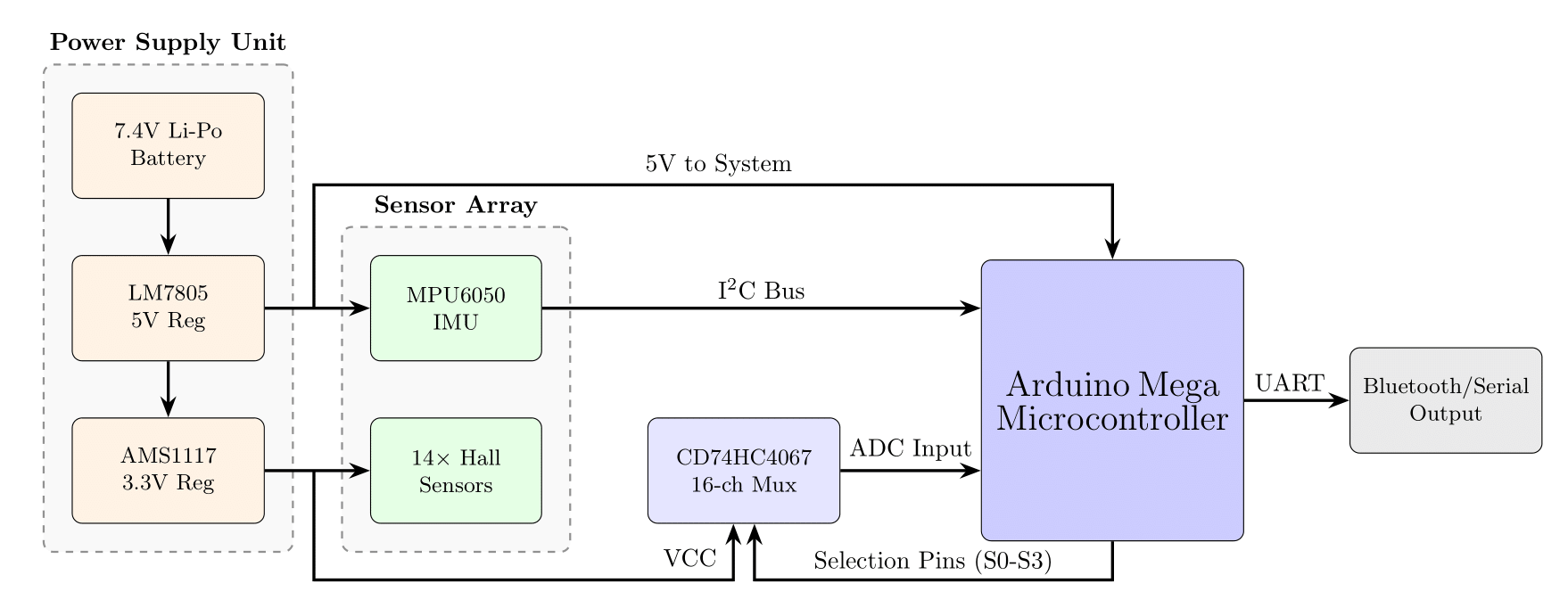}
    \caption{Block diagram of the electronic system architecture}
    \label{fig:circuit_diagram}
\end{figure}

The proof-of-concept prototype uses an Arduino Mega microcontroller for rapid development and validation. This platform provides sufficient analog input pins, processing power, and ease of programming for embedded neural network implementation as shown in the circuit architecture in \ref{fig:circuit_diagram}.

\subsubsection{Power management cascade.} Stable analog sensing demands noise-free supply rails. A two-stage regulator chain therefore conditions the 7.4~V Li-Po battery: an LM7805 linear regulator feeds a regulated 5~V rail for the Arduino $V_{\text{in}}$ pin, and an AMS1117 low-dropout regulator subsequently generates a dedicated 3.3~V rail for the 14 Hall sensors, the MPU6050, and the CD74HC4067 multiplexer. Segregating the sensor rail prevents digital switching noise from injecting artifacts into the analog readings.

\subsubsection{Signal multiplexing and wiring.} To maintain a compact design and allow for future scalability, all 14 Hall-sensor outputs terminate on a CD74HC4067 16-channel analog multiplexer. The individual sensor leads occupy inputs C0--C13, while the shared output (SIG) connects to one of the Arduino's analog input pins. Four digital pins drive the S0--S3 selection lines, stepping through binary combinations (0000--1111) so that each joint is sampled sequentially with minimal pin usage.

\subsubsection{Inertial measurement augmentation.} The system utilizes an MPU6050 module connected over I2C to supply essential accelerometer and gyroscope readings. This data allows the model to differentiate between gestures with similar hand shapes by analyzing wrist motion and orientation, effectively separating, for example, a vertical pointing gesture from a forward one.

\section{Firmware and Software Ecosystem}
The computational workflow spans three cooperating layers: embedded firmware on the glove, an offline training environment that prepares neural weights, and application-level interfaces for users.

\subsection{Embedded Firmware}
The C++ firmware running on the Arduino Mega cycles between a data-collection state and an inference state that performs real-time classification. Sampling, preprocessing, inference, and communication are organized as a deterministic loop, ensuring bounded latency for interactive use.

\subsubsection{Operational feedback.} An RGB LED mounted on the enclosure exposes the firmware state without requiring a display. Red indicates boot, yellow denotes serial communication initialization, magenta flags MPU6050 initialization faults, blue signals that the inertial unit is calibrated while awaiting data collection, and green confirms active inference mode.

\subsubsection{Neural inference engine.} Gesture recognition is executed on-device by a Multi-Layer Perceptron implemented in C++. The topology mirrors the trained model, comprising an input layer equal to the normalized sensor vector length (20 inputs: 14 Hall-sensor channels and 6 IMU channels), a single sigmoid-activated hidden layer, and a sigmoid output layer corresponding to the gesture vocabulary. Each inference cycle aggregates the 14 Hall-sensor voltage readings and six IMU channels, normalizes all values to the $[0,1]$ interval, and performs forward propagation through the network. The ANN learns to map the complex, nonlinear sensor patterns directly to gesture classes without explicit angle calculation. The lightweight matrix operations maintain low latency well within the human perceptual threshold ($<$100ms), delivering class probabilities for real-time gesture recognition.

\subsection{AI Training Pipeline (Python/TensorFlow)}
\begin{figure}[htbp]
    \centering
    \includegraphics[width=0.48\textwidth]{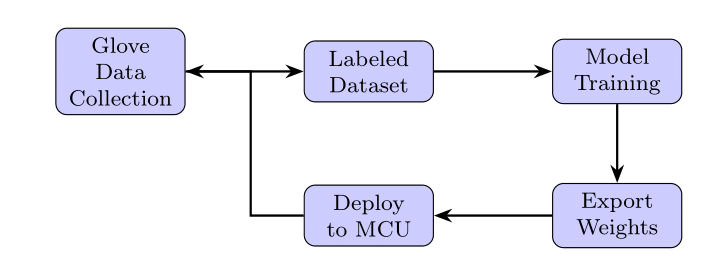}
    \caption{Neural network training workflow from data collection to on-device deployment}
    \label{fig:ai_pipeline}
\end{figure}

Because microcontrollers lack the computational resources to train neural networks efficiently, data-driven learning occurs on a host computer and only the trained weights are deployed to the glove. \ref{fig:ai_pipeline} shows the training pipeline used to train and deploy the ANN model.

\subsubsection{Data collection.} During acquisition sessions, the glove streams raw sensor vectors over USB while multiple participants perform the target vocabulary. A Python utility records these streams and packages them with gesture labels into CSV datasets.

\subsubsection{Model training and export.} A TensorFlow/Keras notebook loads the dataset, applies the same normalization used on the microcontroller, and instantiates an architecture identical to the embedded MLP. Training employs backpropagation with an Adam optimizer and categorical cross-entropy loss while monitoring learning curves for overfitting. Once the validation accuracy exceeds the design threshold (96\% for this study), a helper script extracts the learned weights and biases and formats them as C++ arrays.

\subsubsection{Deployment workflow.} The exported arrays are pasted into the firmware source, after which the Arduino is reprogrammed. This manual yet deterministic process ensures the embedded network matches the model verified during offline experimentation.

\subsection{Integration and Application Layer}
The glove interfaces with two complementary user-facing applications. A Unity-based visualization renders a virtual hand driven directly by the sensor stream, facilitating rapid verification of mechanical alignment and signal fidelity. For assistive communication, a Python service running on a paired computer or smartphone listens to the serial or Bluetooth output, translates gesture indices into words, and invokes either Azure Cognitive Services or the offline \texttt{pyttsx3} engine to synthesize speech. Macro-style mappings also permit gestures to trigger keyboard events, enabling auxiliary scenarios such as presentation control.

\section{Results and Discussion}
The prototype was subjected to rigorous testing to validate its design hypotheses regarding cost, accuracy, and usability.

\subsection{Accuracy Assessment}
The system's classification performance was evaluated using a validation set of 11 fundamental Nepali Sign Language words. Data was acquired from five subjects with varying hand anthropometry (25th to 75th percentile) to ensure the model could generalize across users. The on-device ANN achieved an aggregate recognition accuracy of 96\%.

The sensors demonstrated high stability for static postures. While the inclusion of the MPU6050 successfully disambiguated signs with identical hand shapes but differing wrist orientations, some error persistence was noted in gestures requiring fine motor distinction. These results must be contextualized within the physical constraints of the current iteration. The reported accuracy was achieved using a proof-of-concept chassis (Figure~\ref{fig:prototype}) constructed from cardboard. While sufficient for validatiang the Hall-effect sensing principle and the electronic pipeline, this chassis lacks the mechanical rigidity of the final 3D-printed design. The minor flexibility of the cardboard mount introduces mechanical noise that a rigid exoskeleton would eliminate. Therefore, the 96\% accuracy represents a baseline floor; we anticipate that the transition to the rigid 3D-printed structure will further suppress mechanical noise and enhance classification confidence, particularly as the vocabulary scales.

\subsection{Cost-Benefit Analysis}
A detailed Bill of Materials (BOM) analysis confirms the economic viability of the project. The total prototype cost is approximately \$80-\$100, comprising: Arduino Mega microcontroller (\$10-\$15), sensors including 14$\times$ 49E Hall sensors and magnets (\$10-\$15), multiplexer and power ICs (\$2-\$5), glove and 3D printing material (\$10), and battery (\$10). Comparing this to the market landscape reveals a massive disparity. Our multi-purpose Hall-effect sensor glove costs less than \$100 with 96\% accuracy and durable non-contact sensors, while the Bright Sign Glove costs approximately \$3,100 despite using fragile flex sensors. The Cornell Research Glove costs approximately \$900, and SR Robotics costs approximately \$670 with limited vocabulary. The proposed solution is roughly 30 times cheaper than the leading commercial competitor while offering superior durability due to the non-contact nature of the sensors. This cost reduction transforms the device from a luxury item into a scalable commodity that could realistically be distributed by NGOs or government bodies in Nepal.

\subsection{Durability and Performance}
Durability testing highlights the advantage of the Hall-effect approach. In stress tests where flex sensors typically begin to show resistance drift (changing baseline values due to material fatigue), the Hall-effect sensors remained consistent. The magnetic field strength of a permanent magnet does not degrade with mechanical cycle count. The only moving parts are the glove fabric and the 3D printed mounts, which are easily replaceable and non-electronic. This suggests a lifespan significantly longer than resistive alternatives, a crucial factor for a device intended for daily use in resource-constrained settings.

\subsection{Implications and Limitations}
The successful development of this prototype provides a blueprint for solving the accessibility problem in assistive technology. By localizing the design, both in terms of the sign language vocabulary and the supply chain economics, this research offers a tangible solution to the educational and social exclusion of the deaf community in Nepal. The low cost implies that equipping the 22 deaf schools in Nepal with these devices is a financially feasible goal for government or CSR initiatives, potentially revolutionizing classroom interaction.

Despite the success, the prototype faces specific engineering challenges. The MCP joints (knuckles) remain difficult to track perfectly due to their range of motion (flexion plus abduction), making rigid sensor placement difficult. Hall sensors are sensitive to external magnetic fields, and while the local magnets are strong, strong external fields (e.g., from large speakers or motors) could theoretically induce noise. The neural network approach requires training data collection for each new gesture, and while the ANN generalizes well across users, some individual calibration may improve accuracy.

\textbf{Vocabulary scalability:} Mechanical constraints limit the current system to 11 words. The cardboard chassis serves for sensor validation but lacks rigidity. It allows minor sensor displacement during rapid movement. This reduces the precision needed to distinguish similar signs. A rigid 3D-printed exoskeleton solves this issue. It stabilizes the magnetic field baseline. Tighter mechanical tolerances will support a vocabulary of hundreds of signs. Algorithmic changes are also necessary for expansion. The current feed-forward MLP classifies static poses effectively. Fluent signing involves temporal sequences. Future iterations could use Recurrent Neural Networks (RNNs) or Long Short-Term Memory (LSTM) units. These architectures model time-dependent patterns.

\section{Conclusion}
This project validates the Hall-effect sensor glove as a superior alternative to resistive flex sensor designs. We prioritized durability and local economic constraints. The result challenges the assumption that accurate gesture recognition requires expensive hardware. Integrating Hall-effect sensors with an embedded neural network yielded 96\% accuracy. The total cost remained under \$100. This confirms that strategic engineering delivers high-precision results without high-cost components.

The current prototype classifies static poses. Future work could target continuous sign language translation. Transitioning from simple MLP networks to LSTM architectures would enable the system to interpret dynamic gesture sequences. Replacing the prototype chassis with a refined 3D-printed mechanical design and migrating from the Arduino Mega development board to a custom-designed ASIC would reduce size, power consumption, and production costs while enabling more sophisticated neural architectures. These improvements would expand the vocabulary to cover full NSL conversation. Pilot trials in Kathmandu's deaf schools could refine the device based on direct user feedback and enable deployment of a sustainable assistive tool for the Nepali deaf community.

\bibliographystyle{IEEEtran}
\bibliography{ref_rep}

\end{document}